\documentclass[reprint, amsmath, amssymb, aps,prl,floatfix,showpacs,amssymb,unsortedaddress]{revtex4-2}
\usepackage[utf8]{inputenc}
\usepackage{amssymb,amsfonts,amsmath,bm}
\usepackage{graphics,graphicx}
\usepackage{color}
\usepackage{siunitx}
\usepackage{hyperref}

\newcommand{\lB}{{l_\mathrm{B}}}

\usepackage{verbatim}

\begin{document}

\title{
Like-charge attraction at the nanoscale: \\ ground-state correlations and water destructuring
}

\author{Ivan Palaia}
\affiliation{Institute of Science and Technology Austria, 3400 Klosterneuburg, Austria}

\author{Abhay Goyal and Emanuela Del Gado}
\affiliation{Department of Physics, Institute for Soft Matter Synthesis and Metrology,
Georgetown University, 37th and O Streets, N.W., Washington, D.C. 20057, USA}

\author{Ladislav \v{S}amaj}
\affiliation{Institute of Physics, Slovak Academy of Sciences, 84511 Bratislava, Slovakia}

\author{Emmanuel Trizac}
\email{emmanuel.trizac@universite-paris-saclay.fr}
\affiliation{Universit\'e Paris-Saclay, CNRS, LPTMS, 91405, Orsay, France.}

\begin{abstract}
    Like-charge attraction, driven by ionic correlations, challenges our understanding of electrostatics both in soft and hard matter. For two charged planar surfaces confining counterions and water, we prove that even at relatively low correlation strength, the relevant physics is the ground-state one, oblivious of fluctuations. Based on this, we derive a simple and accurate interaction pressure, that fulfills known exact requirements and can be used as an effective potential. We test this equation against implicit-solvent Monte Carlo simulations and against explicit-solvent simulations of cement and several types of clays. We argue that water destructuring under nanometric confinement drastically reduces dielectric screening, enhancing ionic correlations. Our equation of state at reduced permittivity therefore explains the exotic attractive regime reported for these materials, even in absence of multivalent counterions.
    
    
\end{abstract}

\date{\today}

\maketitle

\section*{Introduction}

When two identical charged colloids are immersed in a solvent, their electrostatic interaction is mediated by fluctuating smaller species, such as microions \cite{HansenLowen,Bell00,Hunter,Levin2002,Andelman2010,Zhao2018}. Pioneered by Gouy \cite{Gouy1910} and Chapman \cite{Chapman1913}, the statistical treatment of this phenomenon, accounting for thermal fluctuations, is a cornerstone of colloid science and goes by the name of Poisson-Boltzmann theory \cite{Hunter,Messina2009}. Within such theory,  macromolecules with a bear charge of the same sign invariably experience a repulsive force, which provides the Coulombic contribution to the DLVO theory 
\cite{Hunter,Zhao2018}. 
In this framework, electrostatic interactions 
between similar bodies, of arbitrary geometry, are necessarily repulsive \cite{Neu1999}.
However, as initially shown by Monte Carlo simulations \cite{Guldbrand1984} and integral equations studies \cite{Kjellander1984}, like-charge macromolecules in solution can attract. This counterintuitive phenomenon is the hallmark of electrostatic correlations between ions \cite{Jonsson2004JA}. Experiments and system-specific simulations proved it to be of paramount importance to explain cement cohesion
\cite{Pellenq2004, Pellenq1997, Plassard2005}, docking of vesicles \cite{Komorowski18,Komorowski20}, DNA condensation in viruses or cells \cite{Bloomfield1991}, 
as well as the behavior of like-charged mica surfaces \cite{Kekicheff93}, polyelectrolytes \cite{Angelini}, lamellar systems \cite{Khan1985}, and lipid bilayers \cite{Mukina19,Fink19}. A time-honored rule of thumb is that like-charge attraction requires multivalent counterions
\cite{Bell00,Levin2002,Messina2009}.

From a theoretical standpoint, like-charge attraction provides
a complex many-body problem \cite{RoBl96,Allahyarov,Shkl99,Levin99,Lau,Netz2001,Moreira2002,Chen2006,Santangelo2006,Pegado,Hatlo2009,Bask2011,Naji2013}. Analytical progress is solely possible in the case where counterions are the only small species present (no added salt) and for simple geometries, e.g.~where point ions are confined
in water between two planar charged surfaces. These simplifications maintain physical relevance: on the one hand, confinement often leads to coion exclusion and no-salt conditions \cite{Yang,Hishida2017,Tournassat2016}; on the other hand, effective interactions for more complex geometries can be obtained by means of a Derjaguin approximation, once the planar geometry has been solved \cite{Israelachvili}. 

\begin{figure}[htbp]
\begin{center}
\centering
\includegraphics[width=1\columnwidth]{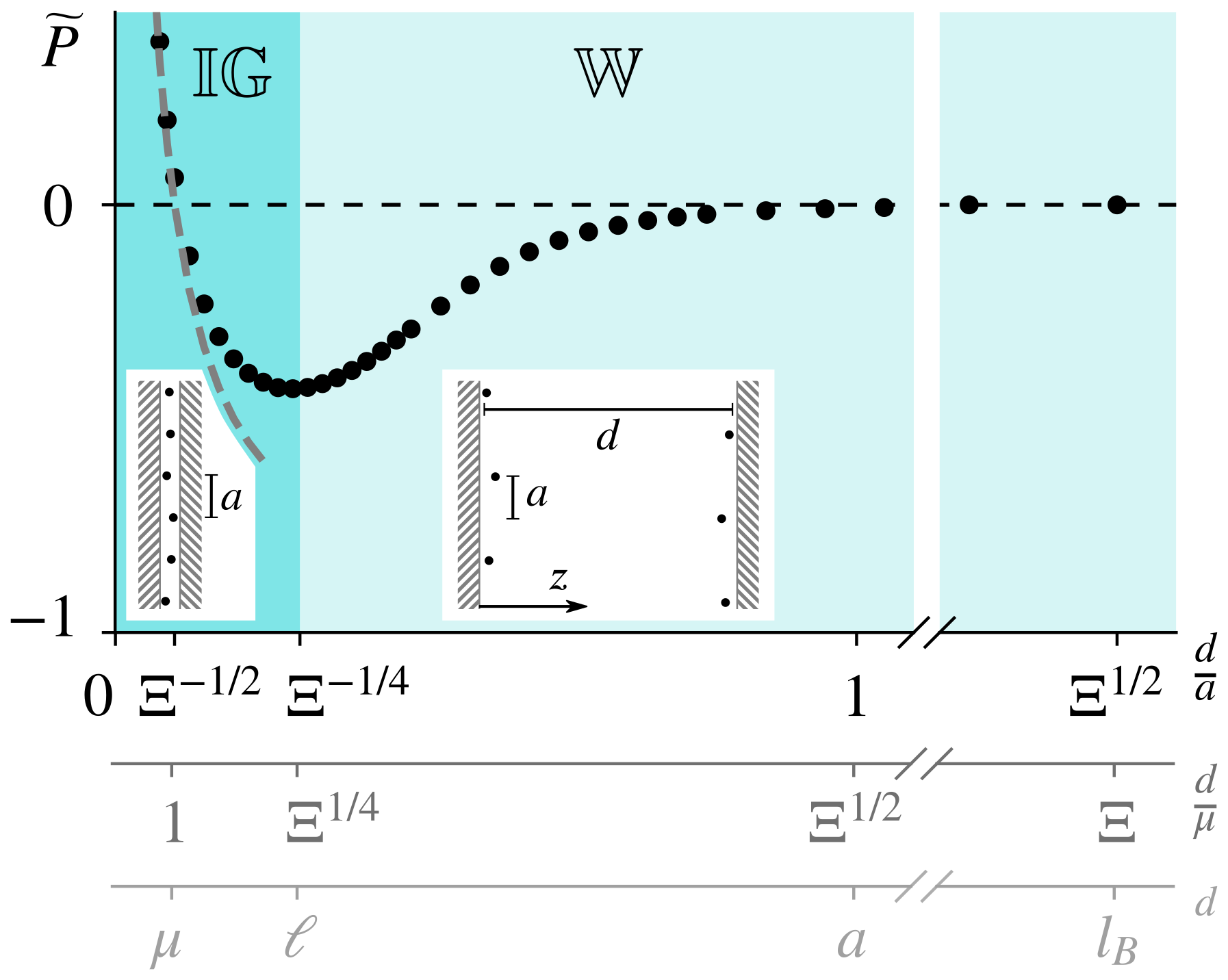}
\caption{Typical pressure profile between two charged plates at distance $d$. Here $\Xi=100$ (Monte Carlo simulations from \cite{Samaj2018a}), but this shape holds whenever an attractive branch exists, meaning $\Xi>12$. The four relevant length scales are specified below the graph (neglecting numerical prefactors) in units of the Gouy-Chapman length $\mu$ and of lateral spacing $a$. Note the scaling, that helps calculations: $q^2 \lB/a \propto a/\mu \propto (\ell/\mu)^2\propto \sqrt\Xi$.
The cartoons in the insets show microion positions in the slab
for the two possible regimes: a short distance ``ideal gas'' region ($\mathbb{IG}$) for $d<\ell$ where Eq.~\eqref{eq:ig} applies (shown by the gray dashed line), and a Wigner region ($\mathbb{W}$) for $d>\ell$. 
The system is globally electroneutral.
}
\label{fig:sketch}
\end{center}
\end{figure}

We shall address the problem of understanding the equation of state of a correlated salt-free system, i.e.~how the force between charged plates changes with their distance. The system is represented in the insets of Fig.~\ref{fig:sketch}, while the rest of the figure shows a sketch of its  equation of state, which is non-monotonic. 
We argue below that the challenge is to understand the 
increasing $\mathbb{W}$ branch, while the short distance $\mathbb{IG}$ regime follows from a simple ideal gas argument. More importantly, it is the $\mathbb{W}$ branch that is relevant for a number of applications.  While much analytical and computational 
effort has been invested in this very question and the role of ionic correlations emphasized
\cite{RoBl96,Shkl99,Netz2001,Moreira2002,Chen2006,Santangelo2006,Hatlo2009,Kanduc2008,Bask2011,Naji2013,Samaj2018a,Gronbech97,LiLo99,Jonsson2004JA,Pegado},
most theories fail at accounting for the $\mathbb{W}$ branch.

Our motivation is threefold. First, we show that the 
$\mathbb{W}$ branch is more universal than previously thought and is closely related to the zero-temperature equation of state.
Second,
elaborating on a number of exact results, we derive a versatile and accurate equation for the pressure, that covers not only the 
$\mathbb{W}$ branch, but the whole distance range. 
This equation of state passes the tests of exact known limiting behaviors whilst remaining simple, at variance 
with previous attempts. The third objective of this paper is to investigate a situation where our slab setting is of much relevance, that is under extreme confinement (small interplate distance $d$, on the order of the size of a water molecule), where the molecular nature of the solvent cannot be disregarded. In line with a number of recent works \cite{Schlaich2016,Fumagall2018,Schlaich2019,Loche2020,Mukina19,cement}, it then becomes essential to account for the destructuring of the water network in the slab, which leads to a dramatic decrease of screening, and a concomitant increase not only of correlations, but also of the attractive force between the surfaces. We show that our equation of state applies here in a ``vacuum'' reformulation that we coin the ``locked water picture'', and is directly relevant for clays and cement,
where key agents are found in the form of charged nanoplates \cite{Qomi2017,Carrier,cement}. 

\section*{Results and discussion}

\textbf{Relevance of ground-state physics.}
We start by a length scale analysis 
\cite{Netz2001,Naji2013},
that provides a fresh overlook.  
Until stated otherwise, the solvent is considered implicitly by a constant (bulk) permittivity $\varepsilon_r$  relative to that of vacuum
$\varepsilon_0$ ($\varepsilon_r \simeq 80$ for water); microions are pointlike
with valence $q$. This yields the so-called primitive model, where the  Bjerrum length $\lB = e^2 /(4\pi\varepsilon_r\varepsilon_0 kT) $ is
the distance at which thermal energy $kT$ matches the interaction potential between two elementary charges $e$. The two plates are modelled as hard surfaces, bearing
a homogeneous surface charge density $\sigma e$, which defines  a second important length, named after Gouy \cite{Gouy1910} and Chapman \cite{Chapman1913}: $\mu = 1/(2\pi \lB q \sigma )$ is the distance 
that can be reached with an energy budget $kT$, dragging away a single counterion 
initially at contact with an isolated plate. 
The so-called coupling parameter, quantifying the importance of ionic correlations, follows as
\begin{equation}
    \Xi \,=\, \frac{q^2 \lB}{\mu} \,=\, 2 \pi \,\lB^2\, q^3\, \sigma .
\end{equation}
Comparing electrostatic and thermal energies, $\Xi$ can be viewed as
a dimensionless measure of surface 
charge, inverse permittivity or, more formally, inverse temperature. 
$\Xi<1$ is the regime where Poisson-Boltzmann theory prevails and the two plates repel, while for $\Xi>12$ like-charge attraction sets in \cite{Moreira2002}.
Given that most charged natural or synthetic surfaces have $\sigma < 1 \,\hbox{nm}^{-2}$ and that $\lB\simeq 0.7\,$nm in bulk water, monovalent counterions ($q=1$) lead to small couplings ($\Xi<4$) and fall under the Poisson-Boltzmann repulsive phenomenology.
This explains the rule of thumb alluded to above: attraction, if any, requires $q\geq 2$ 
and is not possible with monovalent ions. Because of its cubic dependence on valence $q$, $\Xi$ can reach or significantly exceed a few tens 
in a wealth of experiments. This is the strong coupling regime we are interested in.
Two additional lengths need to be introduced. First, $a=\sqrt{q/\sigma}$ defines the Wigner lattice spacing
in the ground state \cite{Goldoni1996,Lobaskin2007,SamajPRB}, i.e.~ the distance between neighbour ions when $\Xi\to\infty$ and ions crystallize on each plane.
This quantity, which loses relevance in the Poisson-Boltzmann regime, is indicated in Fig.~\ref{fig:sketch}. It remains essential whenever $\Xi$ is not small.
The last player here is the distance $\ell$ that discriminates between the
two regions, $\mathbb{IG}$ and $\mathbb{W}$ in Fig.~\ref{fig:sketch}. 
We will see that $\ell \propto \sqrt{a\mu}$.
For large $\Xi$, the four lengths are in the order $\mu<\ell<a<\lB$ and their ratios only depend on $\Xi$, as indicated by the axes in Fig.~\ref{fig:sketch}.

The two branches in Fig.~\ref{fig:sketch} correspond to opposite limiting situations. The left branch has ideal gas nature \cite{Netz2001} and is simple to explain: the two plates are so close that all microions lie in the same plane, and the inter-ions electric field vanishes by symmetry (as a consequence, the potential
is quadratic with lateral displacement, a result used below). Since the electric field
due to the equal plates also vanishes in the slab, microions become homogeneously 
distributed along the $z$ coordinate perpendicular to the plate 
(see Fig.~\ref{fig:sketch}): their number density reads $n(z) = 2\sigma/(qd)$ by electroneutrality. We then invoke the contact theorem \cite{BlHL79,Andelman2010,Israelachvili}, a general and exact result relating pressure $P$ to ion density at contact $n(0)$:
\begin{equation}
P=kT\left(n(0)-2\pi\lB\sigma^2\right)\,.    
\label{eq:contactth}
\end{equation}
This yields a rescaled pressure
\begin{equation}
    \widetilde P \, \equiv \, \frac{P}{2 \pi \, \lB \sigma^2 k T} \, =\, 
    \frac{2\mu}{d} -1 \,.
    \label{eq:ig}
\end{equation}
This ideal gas equation of state (dashed gray line in Fig.~\ref{fig:sketch}) is in good agreement 
with the pressure measured in the decreasing $\mathbb{IG}$ regime.

\begin{figure}
\begin{center}
\centering
\includegraphics[width=\columnwidth]{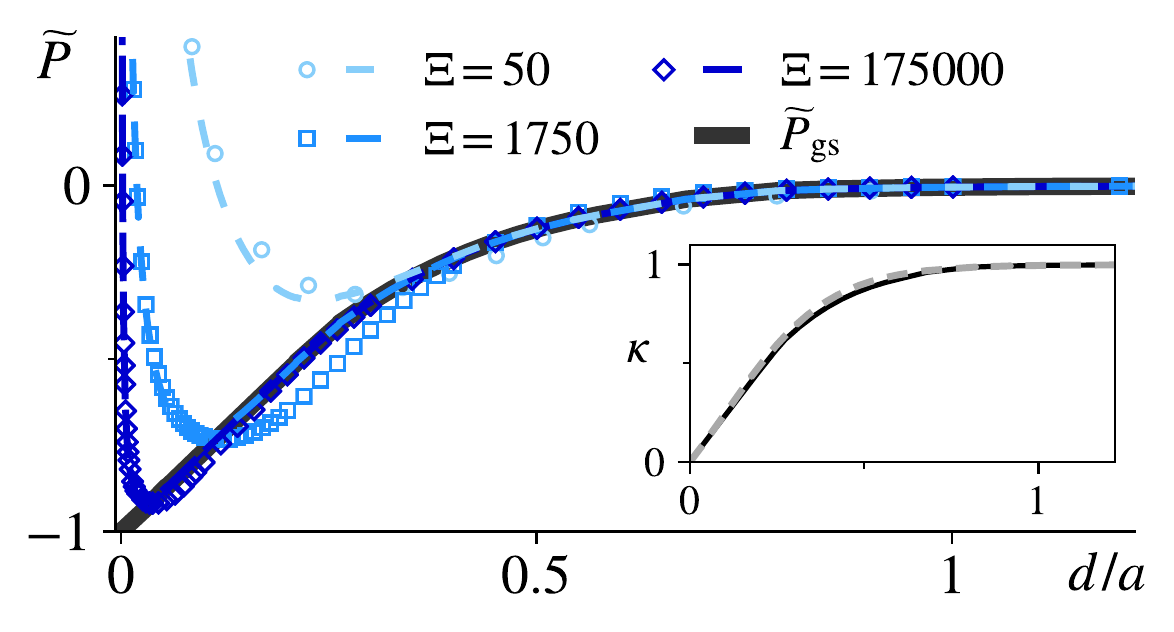}
\caption{Rescaled pressure versus distance. 
Monte Carlo simulations from \cite{Samaj2018a} (symbols) are compared to the ground-state pressure $\widetilde P_{\rm gs}$
($\Xi\to\infty$, continuous black line) and to the analytical equation of state (\ref{eq:Pnew})-(\ref{eq:kappanew}) (dashed lines). 
The inset shows $\kappa$ versus $d/a$: the solid line is
the exact ground-state value \eqref{eq:kap1plusP}, while the dashed line is the analytical approximation \eqref{eq:kappanew}.}
\label{fig:lecrible}
\end{center}
\end{figure}

The complementary $\mathbb{W}$ branch is more subtle, and 
fundamentally many-body. 
We plot in Fig.~\ref{fig:lecrible} the dimensionless pressure $\widetilde P$ as a function 
of $d/a$ for various couplings $\Xi$ (symbols). This reveals a
remarkable collapse: $\widetilde P$, that should converge towards the ground state pressure $\widetilde P_{\rm gs}$ (black line, worked out in \cite{SamajPRB}) only as $\Xi\to \infty$, remains very close to 
this limiting curve down to unexpectedly small $\Xi$.
Even at low coupling, attraction must then stem from the staggering of ions on opposite plates, an ion facing a hole, as in the zero-temperature crystal phase \cite{Goldoni1996,SamajPRB}.
Increasing $d$, one transitions rather abruptly from the ideal gas branch where interactions are immaterial, to the $\mathbb{W}$ branch
that is truly many body. 
The minimum of the pressure curve, giving the crossover
$\mathbb{IG} \leftrightarrow \mathbb{W}$ at $d=\ell$, 
is found by equating both limiting results, Eq.~\eqref{eq:ig} and  $\widetilde P_{\rm gs}$
(we use $\widetilde P_{\rm gs}+1\propto d/a$ at small $d$, as per
Fig.~\ref{fig:lecrible}). This yields 
$\ell \, \propto \, a \,\Xi^{-1/4} \propto \, \mu \,\Xi^{1/4}$. 
This exact result, while confirming findings from other methods \cite{Chen2006}, disproves the so-called Rouzina-Bloomfield criterion \cite{RoBl96,Netz2001}, that had $\ell \propto a$.


\textbf{Analytical equation of state.}
We turn to the second objective of finding an accurate 
equation of state covering both $\mathbb{IG}$ and $\mathbb{W}$
sectors.
Besides ionic positions, the Hamiltonian of the system depends only on the single parameter $\Xi$ \cite{Netz2001}, so that the rescaled pressure has
two arguments: $d$ and $\Xi$. We consider two ways of taking the $\Xi\rightarrow\infty$ limit: by keeping $d/\mu$ constant or by keeping $d/a$ constant. Since $\ell/\mu\propto\Xi^{1/4}\rightarrow \infty$, the former magnifies the $\mathbb{IG}$ branch; conversely, since $\ell/a\propto\Xi^{-1/4}\rightarrow 0$, the latter magnifies the $\mathbb{W}$ branch. 
We can then write the exact relations: 
\begin{align}
& \lim_{\Xi\to\infty} \widetilde P\left(d,\Xi\right)\big\vert_{\frac{d}{\mu}} \, = \,  
\frac{2 \mu}{d}-1 \,;
\label{eq:largeXifixedmu}\\
& \lim_{\Xi\to\infty} \widetilde P\left(d,\Xi\right)\big\vert_{\frac{d}{a}} \, = \, 
\widetilde P_{\rm gs}\left(d\right) \,\,
\label{eq:largeXifixeda}
\end{align}
where $\widetilde P_{\rm gs}$ varies over distances of scale $a$, as shown in Fig.~\ref{fig:lecrible}. In Eq.~\eqref{eq:largeXifixedmu},
the large $\Xi$ limit is taken at fixed $d/\mu$ while it is the ratio
$d/a$ that is held fixed in Eq.~\eqref{eq:largeXifixeda}.
Statement \eqref{eq:largeXifixedmu} indicates a flat, ideal gas profile on a scale $\mu$ (see Eq.~\eqref{eq:ig}). This occurs because the energy cost for an ion to move from one plate to the other is $\Delta E \propto kT \,\lB\, d^2 /a^3 \propto (d/\mu)^2\, \Xi^{-1/2}\rightarrow 0$.
Statement \eqref{eq:largeXifixeda}, instead, expresses the constraint that on a scale $a$ any trace of fluctuations disappears and the ground state pressure is recovered.
We will demand that an approximate pressure
fulfill both constraints, in addition to being trustworthy 
for as wide a range of $\Xi$ values as possible.

To proceed, it is useful to understand which ground-state properties 
are inherited by the finite-$\Xi$ system. 
What is common between infinite and finite (but non-small) $\Xi$ regimes is the local electric field acting on an ion. The reason is that the extent 
of allowed fluctuations along $z$ is always much smaller than $a$.
For an isolated ion on a single plate
(the other plate being at infinite $d$, screened by its own ions), 
this field is attractive, given by $E_0=e \sigma / (2\varepsilon_0 \varepsilon_r )$. In the ground state as well as at finite $\Xi$,
an ion sees similar environments: a layer of ions close to the same plate, that do not create any local field along $z$, and a layer of ions
on the opposite plate, that contribute to renormalize the bare field $E_0$ by a factor $\kappa$, which depends on $d/a$. 
We have just argued that $\kappa \to 1$ for $d/a\to \infty$, while 
$\kappa \to 0$ for $d/a \to 0$ (where one must recover the ideal gas picture). 
More generally, the following simple mechanical argument relates $\kappa$ to $\widetilde P_{\rm gs}$.
In the ground state, each ion at contact with the plate pushes on it with a force $\kappa e q E_0$; there are $\sigma/q$ ions per unit surface, so the repulsive force per unit surface is $\kappa e \sigma E_0$. At the same time, the plate feels an attractive force due to the presence of the ion layer and of the other plate: their overall charge density is $\sigma e$, and so is the charge of the initial plate, modulo a sign, so the force per unit surface is $-(\sigma e)^2/(2\varepsilon_0 \varepsilon_r) = -e\sigma E_0$. The total force acting on the plate is then the sum of these contact and electrostatic components, i.e.\@
 $P_{\rm gs} = (\kappa-1)e\sigma  E_0 = (\kappa-1) \sigma^2 e^2/(2\varepsilon_0\varepsilon_r)$. In dimensionless form:
\begin{equation}
    \kappa
    \,=\, 1 \,+\, \widetilde P_{\rm gs}.
    \label{eq:kap1plusP}
\end{equation}
This relation can be viewed as a contact-value theorem at $T=0$. Exact in the ground state, we will show it is an excellent approximation even at finite $\Xi$. 

Since the ionic layer thickness
is always much smaller than the inter-ion distance $a$ \cite{Samaj2018a}, the ionic profile follows from a single particle argument in the effective potential 
$\kappa q e E_0 z = kT \kappa z/\mu$: thus
$n(z) \propto e^{-\kappa z/\mu}$. We normalize $n$ by imposing electroneutrality (i.e.\@ $\int_0^d\,n(z)\,\mathrm{d}z = 2\sigma/q$) and, using once more the contact theorem \eqref{eq:contactth}, 
 we obtain the equation of state
\begin{equation}
    \widetilde P \,=\, \kappa \  \frac{1+e^{-\kappa d/\mu}}{1-e^{-\kappa d/\mu}} \, -1 \ ,
    \label{eq:Pnew}
\end{equation}
which exhibits a dependence on the two length scales $\mu$,
and $a$ (through $\kappa$). 
Such functional form coincides with the leading 
order of a large-$\Xi$ expansion in the Wigner Strong Coupling approach \cite{Samaj2018a}.
What remains is to find an expression for 
$\kappa=1+\widetilde P_{\rm gs}$. 
It was shown that $\widetilde P_{\rm gs} \sim -3 \exp(- \alpha\, d/a)$ for $d\gg a$,
with $\alpha=4\pi/(3^{1/4} \sqrt{2}) \simeq 6.75$  \cite{SamajPRB}. Given
that $\kappa \to 0$ for $d\ll a$, the simplest form compatible
with the two limits is 
\begin{equation}
    \kappa \,=\, 1 \,-\, \frac{3}{2\, + \, e^{\alpha d /a}} .
    \label{eq:kappanew}
\end{equation}
This is compared with Eq.~\eqref{eq:kap1plusP} in the inset of Fig.~\ref{fig:lecrible}. 

Equation of state  \eqref{eq:Pnew}, supplemented with \eqref{eq:kappanew}, is our jackknife pressure. It
is not exact, but it is the only available simple pressure compatible with exact 
limiting results, such as \eqref{eq:largeXifixedmu} and \eqref{eq:largeXifixeda}. 
It complies with the energetic attraction/entropic repulsion phenomenology put forward in earlier works, e.g.\@ \cite{Jonsson2004JA}.
Fig.~\ref{fig:lecrible} (dashed lines)
illustrates its good accuracy down to $\Xi=50$ when compared to Monte Carlo results (symbols), thus confirming the relevance of ground-state physics at surprisingly low $\Xi$. 

\textbf{Water destructuring under confinement.}
The previous discussion holds within the (solvent implicit) primitive model, where water enters the description only through its permittivitty $\varepsilon_r$. It is customary to take $\varepsilon_r = 80$, the bulk value. Yet, in situations of strong confinement, when $d$ becomes comparable to the size of water molecules, this choice is questionable, if not misguided. Recent works indeed 
point to the fact that water organization is strongly affected at small $d$, with a freezing of orientational degrees of freedom: this decreases its effective $\varepsilon_r$ \cite{Schlaich2019,Mukina19,Fumagall2018,Schlaich2016,Loche2020,cement}. 
One might believe that these considerations ruin implicit-water
approaches, and in particular the primitive model. We show evidence that this is not the case. 

Recently, pure water confined between neutral surfaces below $\SI{2}{\nano\meter}$ was shown to exhibit a relative permittivity between 1 and 4, depending on $d$ \cite{Fumagall2018}. This came as a confirmation of theoretical results that anticipated a drastic decrease in the component of the dielectric tensor perpendicular to the confining surfaces \cite{Schlaich2016,Loche2020}.  
A complete understanding of how confinement and environment affects $\varepsilon_r$ is still lacking, but we speculate that for strongly charged surfaces and in presence of ions, water mobility is suppressed even more strongly, resulting in a permittivity close to that of vacuum.
We argue that decreasing $d$, the number of counterions in the slab remains fixed by electroneutrality, while water content diminishes. Water is then dominantly immobilized in hydration layers
around counterions, and cannot screen Coulombic interactions like it does in the bulk. This behaviour was confirmed in simulations \cite{cement}, where qualitative measurements of both components of the dielectric tensor showed a drastic decrease.
As we seek to understand attraction from first principles only, we make the assumption that water molecules in the slab are so few that none of them is free: this leads to the crude approximation $\varepsilon_r\to1$, for $d$ smaller than a couple nanometres.
We coin the resulting suppressed-permittivity primitive model 
``locked water'' model \cite{cement}: in it, $\lB$ is renormalized, substituted by
$\lB^{\rm locked} = 80\, \lB$ in all expressions. 
Besides dampening van der Waals interactions \cite{Mukina19, Parsegian2005}, this enormously increases coupling to $\Xi^{\rm locked} = 80^2 \, \Xi$ and drives a massive increase in attraction; this is consistent with what proposed in Ref.~\cite{Schlaich2019}, where pressure between confined decanol bilayers was shown to be described by a larger coupling than expected. A consequence is that for small enough $d$, a very small surface charge $\sigma$ becomes sufficient 
to lead to attraction even with monovalent ions ($q=1$), unlike previously thought: the condition for attraction mentioned above, now $\Xi^{\rm locked}>12$, is met for $\sigma q^3 $ just above $ 0.03\,$nm$^{-2}$, at room temperature.

\begin{figure}
\begin{center}
\centering
\includegraphics[width=\columnwidth]{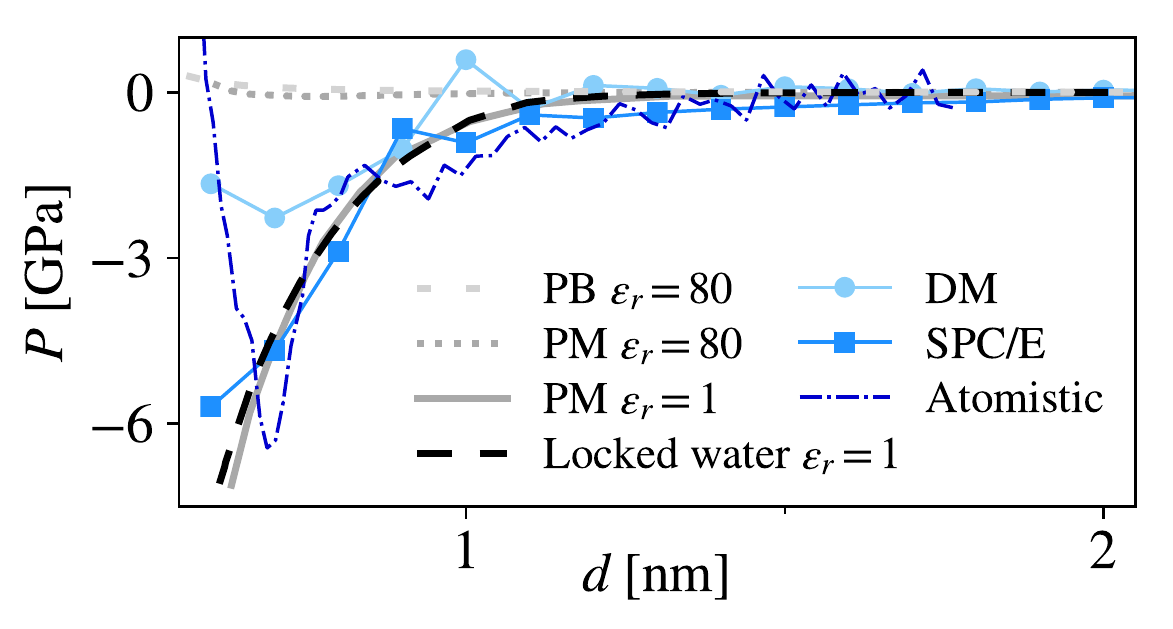}
\caption{Interplate pressure in C-S-H systems, where $\sigma=\SI{3}{\nano\meter^{-2}}$ and ions are Ca$^{++}$.
In blue tones, water-explicit MD simulations from \cite{cement} and \cite{Qomi2017}: dipolar model (DM), SPC/E, and fully atomistic. In gray, MD simulations of the primitive model (PM), at different $\varepsilon_r$, and Poisson-Boltzmann (PB) results. In dashed black, Eq.~\eqref{eq:Pnew}-\eqref{eq:kappanew} at $\varepsilon_r=1$. Curves are shifted horizontally, using SPC/E as reference, to account for different sizes and descriptions of the ions and the plate surface. 
}
\label{fig:cement}
\end{center}
\end{figure}

\begin{figure}
\begin{center}
\centering
\includegraphics[width=\columnwidth]{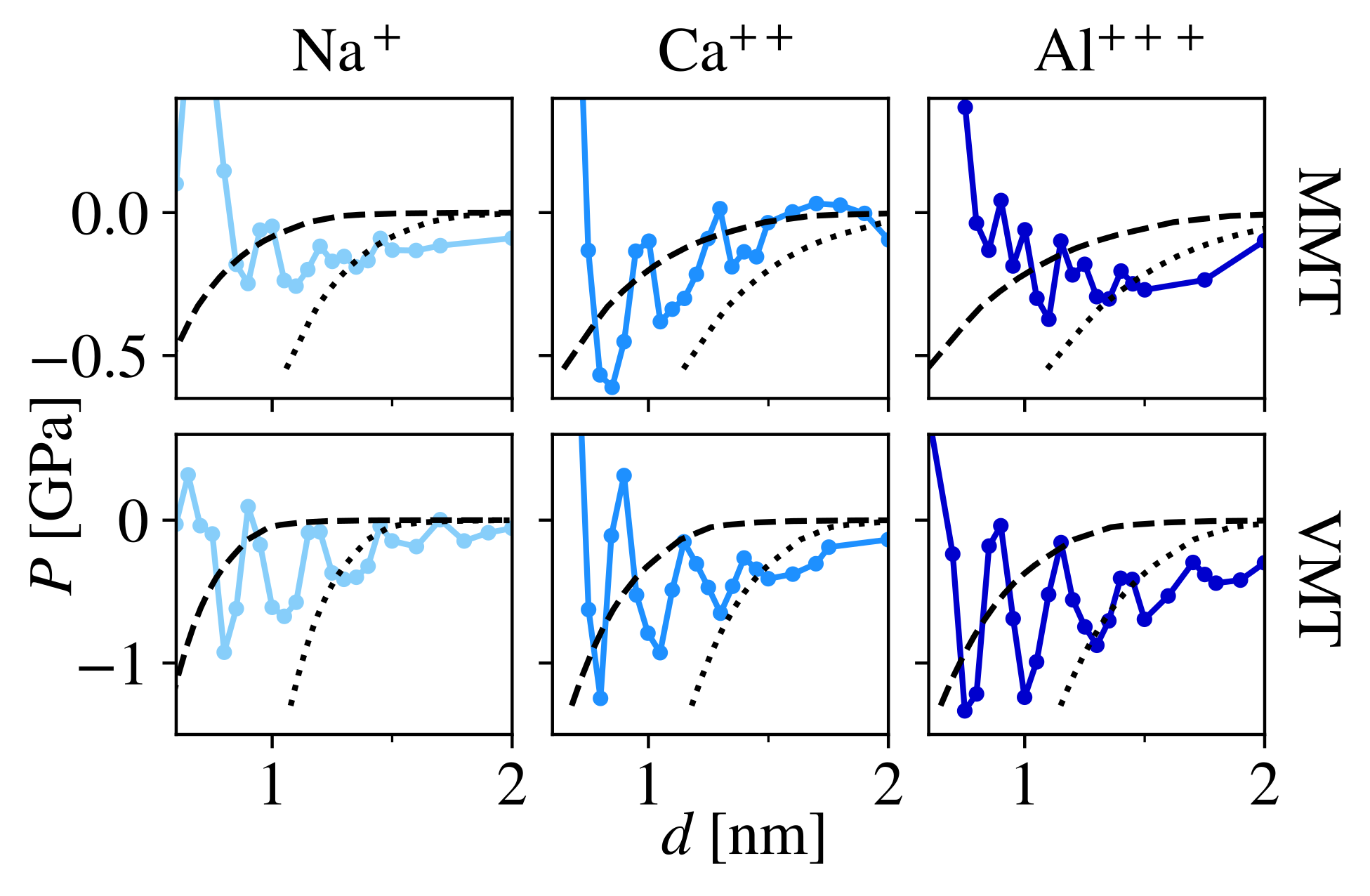}
\caption{Pressure predictions for clays. Above, montmorillonite (MMT, $\sigma=\SI{0.75}{\nano\meter^{-2}}$); below, vermiculite (VMT, $\sigma=\SI{1.5}{\nano\meter^{-2}}$).
Valence $q$ increases columnwise, from 1 to 3. Coupling $\Xi^{\rm locked}$ varies accordingly, from 15000 (top left) to 820000 (bottom right).
Circles show dipolar model MD simulations from \cite{Carrier}, while black lines are from Eq.~\eqref{eq:Pnew}-\eqref{eq:kappanew} at $\varepsilon_r=1$. 
A shift accounts, as in Fig.~\ref{fig:cement}, for finite ion and wall sizes.
In particular, the dotted curve assumes fully hydrated ions, while the dashed one a halved hydration shell.
The ``locked water'' model predicts a pressure between the two black curves.
}
\label{fig:clays}
\end{center}
\end{figure}

The equation of state \eqref{eq:Pnew}-\eqref{eq:kappanew} within the ``locked water'' primitive model can be tested against water-explicit simulations. We take as sample systems cement and clays, whose effective interactions at the nanoscale have puzzled scientists for decades.
Fig.~\ref{fig:cement} compares different models for calcium silicate hydrate (C-S-H), the main binding agent of cement \cite{Pellenq2004}: these include fully atomistic simulations \cite{Qomi2017} and two coarse-grained models \cite{cement}, where 
particles are Lennard-Jones spheres and 
the dielectric properties of a water molecule emerge either from a point dipole (dipolar model, DM) or from point partial charges (SPC/E model \cite{Berendsen1987}). 
Analogously, Fig.~\ref{fig:clays} compares dipolar model simulations of clays from \cite{Carrier} with Eq.~\eqref{eq:Pnew}-\eqref{eq:kappanew} for ``locked water''. In both figures, curves are shifted horizontally to account for different ion sizes and different descriptions of the walls. This shift is not the result of a fit and is determined by parameters used in simulations for wall thickness or ion diameter. Ambiguities resulting from surface roughness or soft potentials are small and have been resolved by checking for the positions of the two furthermost peaks of the counterion density. In addition, while in simulations of cement the high electric fields involved, at any $d$, squeeze counterions against the walls and expel half of their solvation shell \cite{cement}, the lower surface charges of clays are not always sufficient to quench hydration: for clays, the shape of the solvation shells is known to drastically depend on distance, incidentally causing swelling \cite{Hensen2002,Brochard2021}. This is why in Fig.~\ref{fig:clays} we include two analytical curves, shifted two water molecule diameters from one another. The theoretical prediction lies between the two curves. 

Fig.~\ref{fig:cement} shows that Eq.~\eqref{eq:Pnew}-\eqref{eq:kappanew} within our ``locked water'' picture explains results from all three computational models. In particular,
the primitive model at $\varepsilon_r=80$ catastrophically misses the simulated pressure curves by two orders of magnitude, whereas our ``locked water'' picture captures both their magnitude and qualitative behavior. The Poisson-Boltzmann approximation (DLVO) \cite{Andelman2010}, as mentioned, can only predict repulsion \cite{Neu1999} and is completely wrong. 
Fig.~\ref{fig:clays} shows similar results.
Here, the pressure oscillations observed are fingerprints of the water diameter ($\simeq\SI{0.3}{\nano\metre}$) and a full account of these would require a finer
molecular description; nonetheless, general trends and magnitudes are well captured. This is not trivial, as state-of-the-art implicit-solvent theoretical models involve $\varepsilon_r=80$, missing again even the order of magnitude of the pressure.  
Moreover, we explain the observed attraction in presence of monovalent ions (Na$^+$),
recently reported also for lipid bilayers \cite{Mukina19}.

In both Figs.~3 and 4, the dramatic increase from the bulk $\Xi$ to $\Xi^{\rm locked}$ decreases the crossover length $\ell$ by a factor $\sqrt{80}\simeq 9$. This extends the relevance of the $\mathbb{W}$ branch and contributes to masking the $\mathbb{IG}$ regime. In simulations, the steep rise at small distances is indeed not of electrostatic origin, but due to steric ion-wall repulsion. In general, in real systems, the $\mathbb{IG}$ branch and possibly the initial part of the $\mathbb{W}$ branch will be masked by other additive components of the pressure, such as hydration or surface-surface contact forces. These short-range forces, typically described at a phenomenological level \cite{Israelachvili}, are not accounted for in our first-principles model. However, the many experimental and numerical observations of like-charge attraction under confinement, as well as Fig.~\ref{fig:cement} and \ref{fig:clays}, suggest that they are unlikely to screen the longer-range attractive branch of the electrostatic correlation pressure. 

Charge regulation and surface charge heterogeneity are not included explicitly in our model. These aspects might play a role in determining the strength of attraction. For instance, pH is known to drastically regulate charge in cement, as the hydration reaction proceeds \cite{Labbez2006,Labbez2011,cement}. A relevant observation is that, in the ground-state picture that inspires our equation of state, ions are immobilised on the surface, as also observed in simulations \cite{cement}. This is effectively equivalent to having counterions chemically bound to the surface, assuming that they produce a localised excess positive charge on the negative plate. In sum, the only thing that matters to have correlation-induced attraction is a (staggered) non-uniform charge distribution on the plates, be it from localised chemically bound charges or from ions dwelling at the interfaces. Consistently with this scenario, charge regulation with discrete titration sites has been recently suggested to even increase attraction at the nanoscale \cite{Curk2021}. 

\section*{Conclusions}

We have derived a robust, simple and accurate
equation of state for strongly charged plates with ions and water in between. This jackknife pressure, Eq.~\eqref{eq:Pnew}-\eqref{eq:kappanew}, satisfies exact requirements, both at the
scale of the Gouy-Chapman length $\mu$, Eq.~\eqref{eq:largeXifixedmu}, 
and at that of the Wigner spacing $a$, Eq.~\eqref{eq:largeXifixeda}.
This clarifies the rather elusive primitive model phenomenology,
showing that ground state physics, a basic ingredient of our equation of state, is unexpectedly relevant even at moderate coupling.
The origin of attraction must then lie in the fact that ions dwelling next to one plane anticorrelate with those next to the other plane, reminiscent of the staggered lattice formed at zero temperature.  
We have then shown that upon renormalizing the Bjerrum length to consider
its {\em vacuum} counterpart $\lB^{\rm locked}$, it becomes possible, with an implicit-solvent approach, to explain explicit-water results under strong confinement. This ``locked water'' view predicts in particular the possibility of attractive effective interactions with monovalent counterions, at variance with common belief, but explaining the behaviour of clays and lipid bilayers \cite{Mukina19}. Using the present jackknife equation 
of state \eqref{eq:Pnew}-\eqref{eq:kappanew} as an effective potential to upscale coarse-grained simulations is a promising avenue. 

\section*{Acknowledgements}
We thank Martin Trulsson for useful discussions and for providing us with simulation data.  This work has received funding from the European Union’s Horizon 2020 research and innovation programme under the Marie Skłodowska-Curie grant agreement 674979-NANOTRANS. The support received from VEGA Grant No. 2/0092/21 is acknowledged.


\bibliography{./libraryabbr.bib}

\end{document}